\documentclass[10pt,a4,conference]{IEEEtran}
\IEEEoverridecommandlockouts
\usepackage{cite}
\usepackage{amsmath,amssymb,amsfonts}
\usepackage{algorithmic}
\usepackage{graphicx}
\usepackage{textcomp}
\usepackage{xcolor}
\usepackage{caption}
\usepackage{subcaption}
\usepackage{makecell}
\usepackage{multirow}
\usepackage{comment}

\usepackage{caption}
\usepackage{colortbl}
\usepackage{color}
\definecolor{light_grey}{rgb}{0.8,0.8,0.8}
\usepackage{dblfloatfix}

\makeatletter 
\newcommand{\linebreakand}{%
  \end{@IEEEauthorhalign}
  \hfill\mbox{}\par
  \mbox{}\hfill\begin{@IEEEauthorhalign}
}
\makeatother 

\def\BibTeX{{\rm B\kern-.05em{\sc i\kern-.025em b}\kern-.08em
    T\kern-.1667em\lower.7ex\hbox{E}\kern-.125emX}}

\IEEEoverridecommandlockouts
\usepackage{tikz}
\usepackage{textcomp}
\usepackage[hidelinks]{hyperref}
\usepackage{lipsum}
\newcommand\copyrighttext{%
  \footnotesize \textcopyright 2024 IEEE. Personal use of this material is permitted.  Permission from IEEE must be obtained for all other uses, in any current or future  media, including reprinting/republishing this material for advertising or promotional  purposes, creating new collective works, for resale or redistribution to servers or  lists, or reuse of any copyrighted component of this work in other works.
  DOI: \href{<http://tex.stackexchange.com>}{10.1109/GLOBECOM52923.2024.10901634}
  }
\newcommand\copyrightnotice{%
\begin{tikzpicture}[remember picture,overlay]
\node[anchor=south,yshift=10pt] at (current page.south) {\fbox{\parbox{\dimexpr\textwidth-\fboxsep-\fboxrule\relax}{\copyrighttext}}};
\end{tikzpicture}%
}

\begin{document}

\title{On Using Curved Mirrors to Decrease\\ Shadowing in VLC
\thanks{Borja Genoves Guzman has received funding from the European Union under the Marie Skłodowska-Curie grant agreement No 101061853.}
}




\author{
\IEEEauthorblockN{Borja Genoves Guzman\IEEEauthorrefmark{1}\IEEEauthorrefmark{2},
Ana Garcia Armada\IEEEauthorrefmark{2}, and
Maïté Brandt-Pearce\IEEEauthorrefmark{1}}
\IEEEauthorblockA{\IEEEauthorrefmark{1}Electrical and Computer Engineering Dept.,
University of Virginia, Charlottesville, VA 22904 USA}
\IEEEauthorblockA{\IEEEauthorrefmark{2}Signal Theory and Communications Dept., University Carlos III of Madrid, Leganes, Madrid 28911 Spain}
\IEEEauthorblockA{E-mails: bgenoves@virginia.edu, agarcia@tsc.uc3m.es, mb-p@virginia.edu}
}

\maketitle

\copyrightnotice 
\vspace{-\baselineskip}

\begin{abstract}
Visible light communication (VLC) complements radio frequency in indoor environments with large wireless data traffic. However, VLC is hindered by dramatic path losses when an opaque object is interposed between the transmitter and the receiver. Prior works propose the use of plane mirrors as optical reconfigurable intelligent surfaces (ORISs) to enhance communications through non-line-of-sight links. Plane mirrors rely on their orientation to forward the light to the target user location, which is challenging to implement in practice. This paper studies the potential of curved mirrors as static reflective surfaces to provide a \emph{broadening specular reflection} 
that increases the signal coverage in mirror-assisted VLC scenarios. We study the behavior of paraboloid and semi-spherical mirrors and derive the irradiance equations. We provide extensive numerical and analytical results and show that curved mirrors, when developed with proper dimensions, may reduce the shadowing probability to zero, while static plane mirrors of the same size have shadowing probabilities larger than 65\%. Furthermore, the signal-to-noise  ratio offered by curved mirrors may suffice to provide connectivity to users deployed in the room even when a line-of-sight link blockage occurs.
\end{abstract}

\begin{IEEEkeywords}
Curved mirrors, line-of-sight (LoS) link blockage, optical reconfigurable intelligent surface (ORIS), reflector, visible light communication (VLC).
\end{IEEEkeywords}

\vspace{-3mm}
\section{Introduction}\label{sec:Intro}
\vspace{-1mm}

Visible light communication (VLC) has been demonstrated to be a complementary technology to alleviate the congestion in the radio frequency bands produced by the massive increase of wireless communication data. VLC retrofits the light-emitting diode (LED) infrastructure, which has been primarily deployed for illumination, to also provide wireless communication services. However, the large path-loss produced in the visible wavelengths when light goes through obstacles makes VLC unreliable, which impacts its quality of service and limits its mass market adoption.

Prior works have strengthened the VLC technology by proposing cooperative multi-point transmissions~\cite{CoMP_VLC_Conf, CoMP_VLC_Journal}, exploiting reflections coming from the wall~\cite{WallReflectionsIEEEAccess}, or creating angle diversity receivers that allow receiving signals from multiple angles~\cite{ADR_Vehicular}. However, these  techniques involve complex deployments to provide high-quality VLC-based services. Recently, the research community has considered the use of optical reconfigurable intelligent surfaces (ORISs) to make the non-line-of-sight (NLoS) VLC signals stronger. Mirrors have been demonstrated to be the most promising ORIS material, showing a larger maturity and providing a better performance than metasurfaces~\cite{MirrorVSMetasurface}. Mirrors have been explored in ORIS-assisted VLC systems to optimize their orientation and LED-user association with the aim of maximizing the data rate~\cite{IRSIndoorVLC, SumRateMirrors}, the spectral efficiency~\cite{JointResourceMirrors} or the secrecy rate~\cite{SecrecyRateVLC1, SecrecyRateVLC2}. Recently, the authors studied the optimal location of both static and movable mirrors in~\cite{Mirror_Globecom2023, guzman2024resource} to minimize the outage probability, which is one of the main problems in VLC. All these works consider the use of plane mirrors that are optimally oriented so that the light coming from an LED hits in the center of such mirror and it is forwarded to the destination. Indirectly, the light impinging onto the mirror does not only contribute to power level at the desired destination, but it can potentially provide coverage to users located surrounding the destination.

\begin{figure}[t]
         \centering
         \includegraphics[width=\columnwidth]{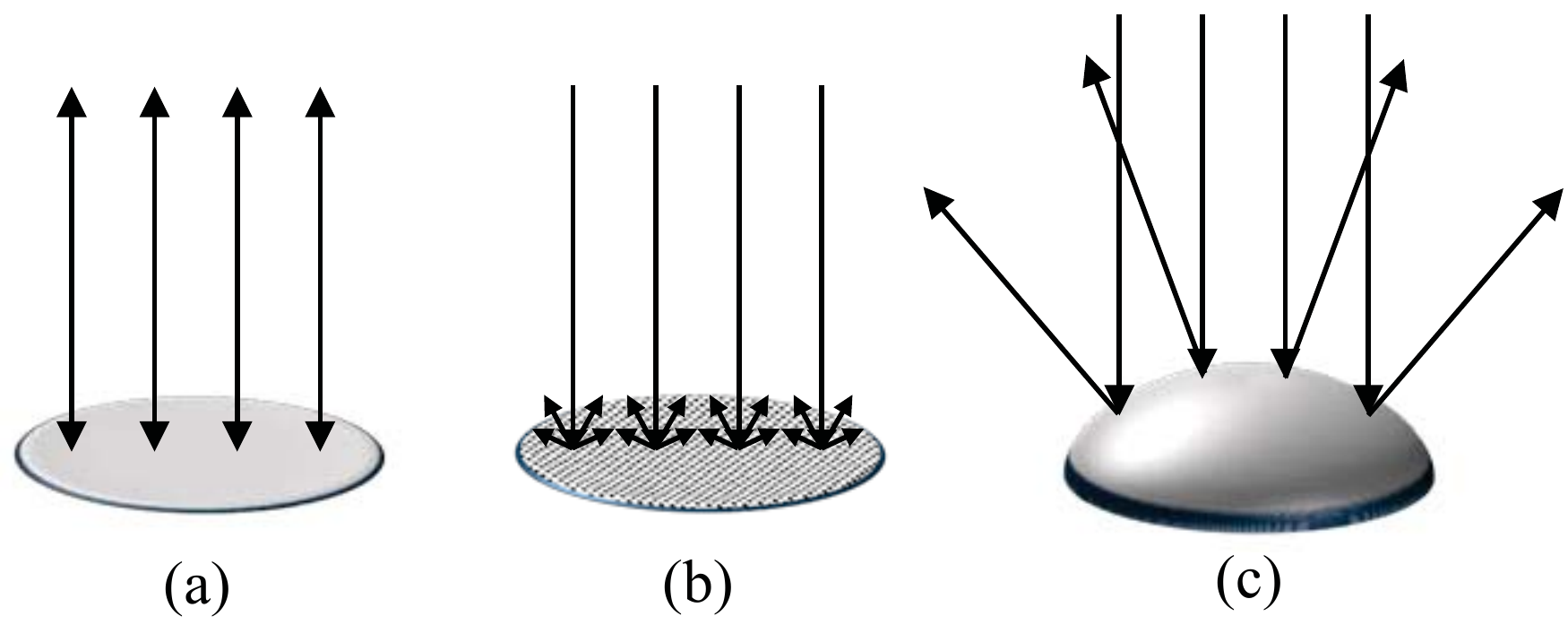}
        \caption{Reflective surfaces: (a) plane mirror (specular reflection), (b) plane wall (diffuse reflection), (c) proposed convex mirror (\emph{broadening specular reflection}
        ).}
        \label{fig:PlaneMirrorAndWallVSCurvedMirrors}
        \vspace{-5mm}
\end{figure}

Although the installation of static mirrors is more realistic than mirrors with optimal and movable orientations, the overall coverage provided by them is still unexplored. 
In the case of a plane mirror, the volume occupied in the space by the rays impinging onto a single reflecting point is the same as the one occupied by the reflected rays. This limits the potential of plane mirrors in VLC when the objective is to provide a coverage as wide as possible. Differently, diffuse reflections produced by walls scatter light, which potentially increases the coverage, but the power loss produced in the material is dramatic. In this paper we introduce the concept of \emph{broadening specular reflection}, 
created by exploiting curved (convex) mirrors where the power loss occurs in the propagation and not in the reflecting material. This phenomenon is depicted in Fig.\,\ref{fig:PlaneMirrorAndWallVSCurvedMirrors}. This paper, for the first time to the best of the authors' knowledge, proposes the use of curved mirrors to increase the coverage in a mirror-assisted VLC system.

We can find few works in the literature that invoke curved mirrors for light communication: Spherical concave mirrors have been proposed in free space optical (FSO) communications to compensate the effects of the air turbulence~\cite{ConcaveMirrorsFSO1, ConcaveMirrorsFSO2};  \cite{CCRMirror} proposed paraboloid mirrors to fabricate a small corner cube retroreflector and provide sunlight communication; spherical reflectors have been installed in user equipment to enable visible light positioning with a broad coverage~\cite{SphereMirrorVLP}; and the authors of~\cite{ConcaveMirrors} theoretically proposed the use of concave mirrors on the transmitter side to expand the lighting area of VLC. Besides, curved convex mirrors have been traditionally placed in corners with low visibility such as hospitals or roads to increase the vision area. Inspired by this traditional use case and different from prior works, we propose the use of convex mirrors placed on walls to extend the VLC coverage and evaluate its performance. 

\emph{Notation:} In this paper, capitalized bold letters such as $\mathbf{A}$ stand for vectors starting at the origin and ending at point A. $\widehat{\mathbf{AB}}$ is the unit vector of $\mathbf{AB}=\mathbf{B}-\mathbf{A}$. $\mathbf{e}_k$ stands for the $k$-th column vector of the 3$\times$3 identity matrix, and $(\cdot)^{\rm T}$ represents the transpose operator.

\vspace{-2mm}
\section{System model}\label{sec:SystemModel}
\begin{figure}[t]
     \centering
         \includegraphics[width=0.7\columnwidth]{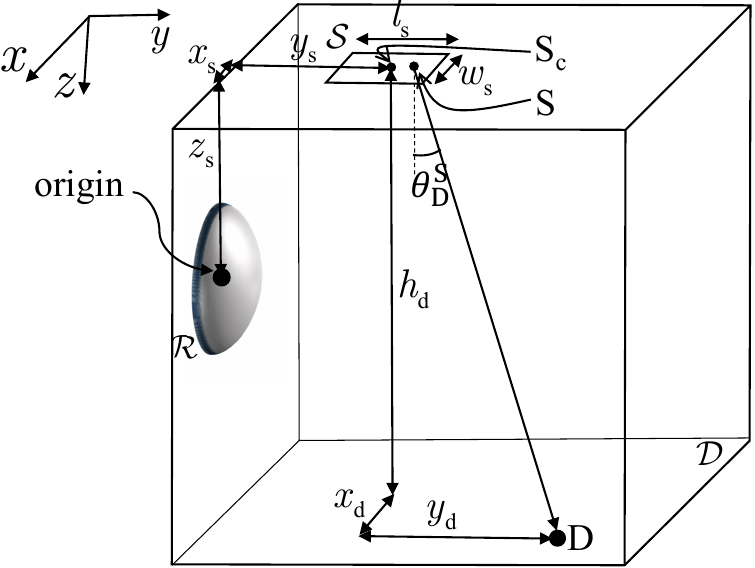}
        \caption{System model including a large source, a point detector, and a convex mirror.}
        \label{fig:SystemModel}
        \vspace{-5mm}
\end{figure}

\vspace{-1mm}
The considered indoor VLC scenario has one light source located at the room ceiling with a height of $h_{\rm d}$ from the horizontal plane where the receiver D is located. The set of points representing the light source and all possible positions of D are denoted by $\mathcal{S}$ and $\mathcal{D}$, respectively. We assume that the source has a uniform power distribution along its area $A_{\rm s}=w_{\rm s}l_{\rm s}$, and that every point in $\mathcal{S}$, denoted by S, has a Lambertian emission pattern. We consider a convex mirror reflector located on a wall, where the set of its points is denoted by $\mathcal{R}$ and whose center is defined as the origin. The detector D is located in a horizontal plane that is parallel to the room floor, and its size is assumed to be a point due to its small active area, typically in the range of mm$^2$ or few cm$^2$. The $x$, $y$ and $z$ axes create a coordinate system with their positive directions as the ones described in Fig.\,\ref{fig:SystemModel}. The point vectors to the center of the source and detector are \mbox{denoted by}
\begin{equation}
\vspace{-2mm}
{\rm \mathbf{S}}_{\rm c} =  \begin{bmatrix}
           -x_{\rm s} \\
           y_{\rm s} \\
           -z_{\rm s}
         \end{bmatrix}, \quad {\rm \mathbf{D}} =  \begin{bmatrix}
           x_{\rm d}-x_{\rm s} \\
           y_{\rm d}+y_{\rm s} \\
           h_{\rm d}-z_{\rm s}
         \end{bmatrix}.
\label{eq:S_D}
\end{equation}

\subsection{Line-of-sight received power}
The lighting industry often installs large lighting sources to provide a homogeneous illumination. Typically, the research community has based its LiFi studies assuming a point source, which puts aside large lighting sources. However, the source size has an important impact on the line-of-sight (LoS) link blockage, and, since large lighting sources are widespread, we consider them in this study.

The total source size is divided into small blocks of size $d$S that transmit an optical power of $P_{\rm opt,S} = P_{\rm opt}/A_{\rm s}$. Assuming that the detector size is very small with respect to the source image in the detector plane~\cite{MirrorVSMetasurface}, the optical power received by the detector located at D through the LoS link is computed as
\vspace{-1mm}
\begin{equation}
\vspace{-1mm}
P^{\mathcal S}_{\rm opt,rx} =  E_{\rm D}^{\mathcal S}\cdot A_{\rm pd},
\label{eq:PoptRxLoS}
\end{equation}
where $A_{\rm pd}$ is the area of the photodetector (PD) and $E_{\rm D}^{\mathcal S}$ is the irradiance at a point D from the source $\mathcal{S}$ that, if detector D is assumed to be looking upwards, can be calculated by
\vspace{-1mm}
\begin{equation}
\vspace{-1mm}
E_{\rm D}^{\mathcal S} = \frac{(m+1)\cdot P_{\rm opt,S}}{2\pi} \int_\mathcal{S} \frac{\cos^{(m+1)}(\theta^{{\rm S}}_{\rm D})}{D^2_{{\rm S,D}}}d{\mathcal S},
\label{eq:IrradianceLoS}
\end{equation}
where $\cos(\theta^{{\rm S}}_{\rm D})=\mathbf{e}_3^{\rm T}\widehat{{\mathbf{SD}}}$, $m=-1/\log_2\left(\cos\left(\phi_{1/2}\right)\right)$ is the Lambertian index of the LED that models the radiation pattern defined by its half-power semi-angle $\phi_{1/2}$, and $D_{{\rm S,D}}$ is the Euclidean distance between S and D points.

\begin{figure}[t]
     \centering
     \begin{subfigure}[b]{0.32\columnwidth}
         \centering
         \includegraphics[width=\textwidth]{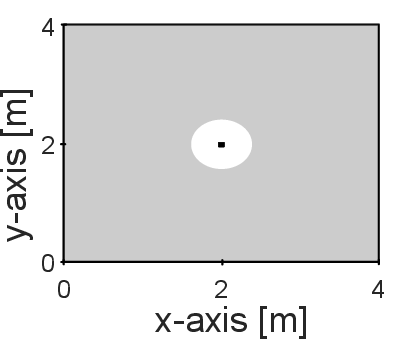}
         \caption{}
         \label{fig:2cm2cm}
     \end{subfigure}
     \hfill
     \begin{subfigure}[b]{0.32\columnwidth}
         \centering
         \includegraphics[width=\textwidth]{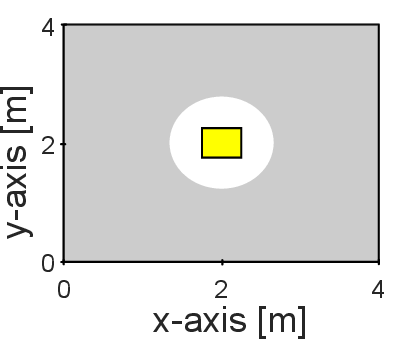}
         \caption{}
         \label{fig:20cm20cm}
     \end{subfigure}
          \hfill
  \begin{subfigure}[b]{0.32\columnwidth}
         \centering
         \includegraphics[width=\textwidth]{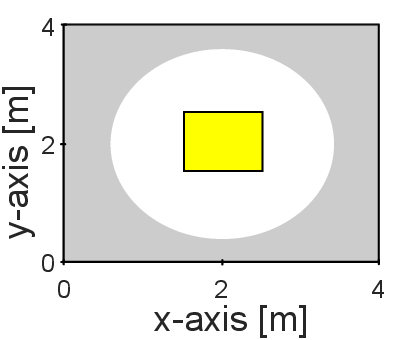}
         \caption{}
         \label{fig:1m1m}
     \end{subfigure}
        \caption{Impact of source size in the LoS-link blockage. Shadowed area represents locations in which the user can not potentially receive a LoS link from the source, and the rectangle located in the middle of the room represents the light source. Light source sizes are: (a) 2\,cm x 2\,cm, (b) 50\,cm x 50\,cm and (c) 1\,m x 1\,m.}
        \label{fig:LoSlinkBlackageVersusSourceSize}
        \vspace{-6mm}
\end{figure}

We can now evaluate the impact of the source size on the LoS-link blockage probability. The irradiance at D can be either 0 or $E_{\rm D}^{\mathcal{S}}$, depending on whether the detector has a LoS-link blocked or not, respectively. Let us assume a user whose position is uniformly distributed along the room, oriented at any horizontal angle following a uniform distribution $\mathcal{U}[0, 2\pi)$, and assuming that the device is always looking upwards. The user body is modelled by a cylinder of height 1.75\,m and radius 0.15\,m, with the device separated at a distance of 0.3\,m from the body~\cite{CylinderBlockage}. The room height is 3\,m, $h_{\rm d}=2$\,m, and the room and source sizes and location are the ones depicted in Fig.\,\ref{fig:LoSlinkBlackageVersusSourceSize}. After distributing $10^4$ users, we compute the user locations where the LoS-link with all points in $\mathcal{S}$ may be potentially blocked. In Fig.\,\ref{fig:LoSlinkBlackageVersusSourceSize} we represent those user locations with a grey shadowed area. Note that the area changes with respect to the size of the source. The larger the source size, the smaller the area representing those locations suffering from LoS-link blockage. The user locations that most likely suffer from a LoS-link blockage are those close to the edges, and they must rely on reflections coming from the walls. This issue was also pointed out by the authors in~\cite{guzman2024resource, Mirror_Globecom2023}, and it was addressed by looking for the optimal location and orientation of planar mirrors. Instead, in this work we aim at simplifying the scenario and analyzing the performance of NLoS coverage when employing a convex mirror that provides a \emph{broadening specular reflection} 
to cover those user locations at the edges.

\vspace{-1mm}
\subsection{Non-line-of-sight received power}
\vspace{-1mm}

\begin{figure}[t]
     \centering
     \begin{subfigure}[b]{0.49\columnwidth}
         \centering
         \includegraphics[width=0.7\textwidth]{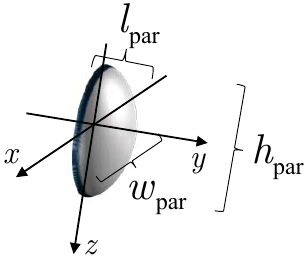}
         \caption{}
         \label{fig:Paraboloid}
     \end{subfigure}
     \hfill
     \begin{subfigure}[b]{0.49\columnwidth}
         \centering
         \includegraphics[width=0.7\textwidth]{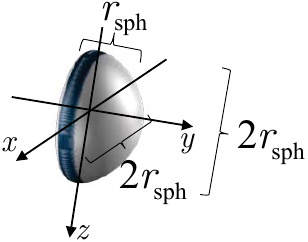}
         \caption{}
         \label{fig:Sphere}
     \end{subfigure}
        \caption{Curved mirrors evaluated: (a) paraboloid and (b) semi-sphere.}        
        \label{fig:CurvedMirrors}
        \vspace{-6mm}
\end{figure}

For the sake of simplicity, we analyze the performance of two three-dimensional reflectors. Namely, we study an elliptic paraboloid and a semi-sphere. Both are depicted in Fig.\,\ref{fig:CurvedMirrors}. The former is defined by the parameters $w_{\rm par}$, $l_{\rm par}$ and $h_{\rm par}$, which indicate the dimensions in the $x$, $y$ and $z$ axes, respectively.  The latter is only defined by the parameter $r_{\rm sph}$, which indicates the radius of the sphere. 
The set of points $\mathcal{R}$ can be either $\mathcal{R}_{\rm par}$ or $\mathcal{R}_{\rm sph}$ when invoking a paraboloid or a semi-sphere, respectively. For simplification, let us use the term $\mathcal{R}$ to refer to the reflective surface in a general way, without specifying its shape.

The optical power received by the detector located at D through the NLoS reflector links is computed as
\vspace{-1mm}
\begin{equation}
\vspace{-2mm}
P^{\mathcal R}_{\rm opt,rx} =  E_{\rm D}^{\mathcal R}\cdot A_{\rm pd},
\label{eq:PoptRxNLoS}
\end{equation}
where $E_{\rm D}^{\mathcal R}$ is the irradiance at a point D from the reflector $\mathcal{R}$, and is derived in Section\,\ref{sec:IrradiancePerformance}.

\vspace{-1mm}
\subsection{Figures of merit}
Two main figures of merit are considered in this study:
\subsubsection{NLoS shadowing probability} To strengthen the VLC performance, we desire to provide a NLoS link to all users distributed along the room that can perform as a backup link in case the LoS one is blocked. This metric defines the probability for a user, uniformly distributed along the room, to have no NLoS contributions due to the incapacity of the reflector. It is represented by 
\begin{equation}
\label{eq:ShadowingProb}
{{\rm Prob}}(E_{\rm D}^{\mathcal R}=0\quad \forall \mathrm{D}\in \mathcal{D}).
\end{equation}
\subsubsection{Signal-to-noise power ratio} It is formulated as
\begin{equation}
\label{eq:SNR}
{\rm SNR}=\frac{\left[\eta_{\rm pd}\cdot \left(\mathbb{I}_{\rm LoS}P^{\mathcal S}_{\rm opt,rx} + P^{\mathcal R}_{\rm opt,rx}\right)\right]^2}{N_0B},
    \vspace{-1mm}
\end{equation}
where $\eta_{\rm pd}$ is the PD's responsivity, $\mathbb{I}_{\rm LoS}$ is the binary variable that takes values 0 or 1 when the LoS link is blocked or not, respectively, $N_0$ is the power spectral density of the additive white Gaussian noise (AWGN), and $B$ is the communication bandwidth. 

\begin{figure}[t]
     \centering
         \includegraphics[width=0.7\columnwidth]{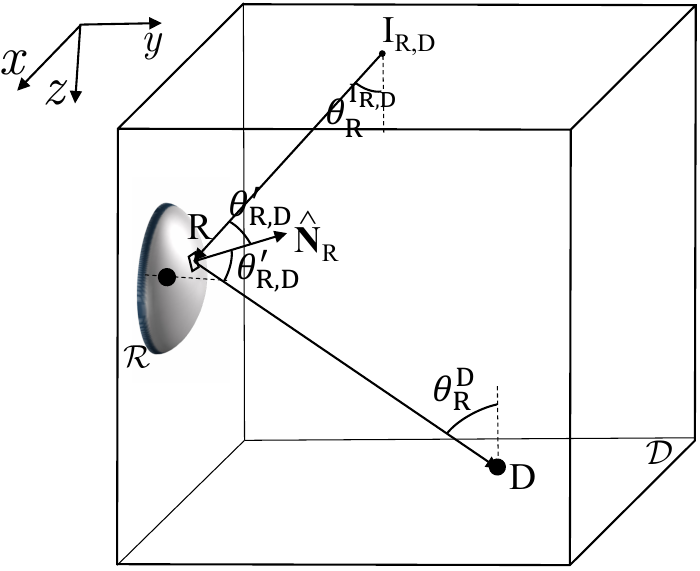}
        \caption{Ray-tracing for a curved-mirror-assisted VLC.}
        \label{fig:AnglesAndVectors}
    \vspace{-4mm}
\end{figure}

\vspace{-1mm}
\section{Irradiance performance of curved mirrors}\label{sec:IrradiancePerformance}
\vspace{-1mm}
The irradiance at a point D from a small point R$\in\mathcal{R}$ can be computed as~\cite[(46)]{MirrorVSMetasurface}
\vspace{-1mm}
\begin{equation}
\vspace{-1mm}
\label{eq:IrradianceAtP}
E_{\rm D}^{{\rm R}}=\rho_{\mathcal R}\cdot L_{\rm R\leftarrow I_{R,D}}\cdot \cos(\theta^{\rm D}_{\rm R})\cdot A_{d\mathcal{R}} \cdot \cos(\theta'_{\rm R,D})/D_{\rm R,D}^2,
\end{equation}
where $\rho_{\mathcal R}$ is the reflection coefficient of the reflective surface $\mathcal{R}$, which is constant for all R points; $L_{\rm R\leftarrow I_{R,D}}$ is the radiance at position R coming from I$_{\rm R,D}$ point; $\theta^{\rm D}_{\rm R}$ is the angle created by the vector $\widehat{\rm \mathbf{DR}}$ and the $z$-axis; $A_{d\mathcal{R}}$ is a small point area of $\mathcal{R}$; $\theta'_{\rm R,D}$ is the incidence angle in R created by the ray coming from I$_{\rm R,D}$ that, due to the Snell's law, is the same as the angle created by the vector $\widehat{\rm \mathbf{RD}}$ and the normal vector of R ($\widehat{\mathbf N}_{\rm R}$); and $D_{\rm R,D}$ is the Euclidean distance between R and D points. The cosine of both angles can be computed by $\cos(\theta^{\rm D}_{\rm R})=\mathbf{e}_3^{\rm T}\widehat{\mathbf{DR}}$ and $\cos(\theta'_{\rm R,D})=\widehat{\mathbf N}_{\rm R}^{\rm T}\widehat{\mathbf{RD}}$. The geometry of the scenario and all its angles are represented in Fig.\,\ref{fig:AnglesAndVectors}. The points R$\in\mathcal{R}_{\rm par}$ can be defined as
\vspace{-2mm}
\begin{IEEEeqnarray}{ll}
\vspace{-1mm}
\label{eq:ParaboloidPoints}
{\rm R}_{\rm par}  {=} 
\bigg\{x,y,z \Big| 
y{=}\frac{x^2}{A}{+}\frac{z^2}{B}{+}C, -\frac{w_{\rm par}}{2}{<}x{<}\frac{w_{\rm par}}{2}, y{>}0,&\IEEEnonumber\\ -\frac{h_{\rm par}}{2}{<}z{<}\frac{h_{\rm par}}{2}\bigg\},
\vspace{-1mm}
\end{IEEEeqnarray}
where $A=-w_{\rm par}^2/(4l_{\rm par})$, $B=-h_{\rm par}^2/(4l_{\rm par})$ and $C=l_{\rm par}$. The points R$\in\mathcal{R}_{\rm sph}$ can be defined as
\vspace{-0.05in}
\begin{IEEEeqnarray}{ll}
\label{eq:SpherePoints}
{\rm R}_{\rm sph} {=} 
\bigg\{x,y,z \Big|
x^2{+}y^2{+}z^2{=}r_{\rm sph}^2, -r_{\rm sph}{<}x{<}r_{\rm sph}, y{>}0,&\IEEEnonumber\\ \vspace{-1mm}
-r_{\rm sph}{<}z{<}r_{\rm sph}\bigg\}. 
\end{IEEEeqnarray}

Differently from works considering a plane reflector where all reflector points have the same normal vector, when considering a curved mirror each point R has its own normal vector, which is computed as
\vspace{-2mm}
\begin{equation}
\vspace{-2mm}
\label{eq:NormalParaboloidVectors}
\widehat{\mathbf{N}}_{{\rm R_{par}}}=\left[\frac{-2x}{A},1,\frac{-2z}{B}\right]^{\rm T}\bigg/\sqrt{\frac{4x^2}{A^2}+1+\frac{4z^2}{B^2}},
\end{equation}
in the case of a paraboloid, and
\begin{equation}
\label{eq:NormalSphereVectors}
\widehat{\mathbf{N}}_{{\rm R_{sph}}}=\left[2x,2y,2z\right]^{\rm T}\Big/\sqrt{4x^2+4y^2+4z^2}
\vspace{-1mm}
\end{equation}
in the case of a semi-sphere. To generalize, we will use the term $\widehat{\mathbf{N}}_{{\rm R}}$.

Since each point in the transmitter is assumed to follow a Lambertian emission pattern, the radiance at position R coming from the I$_{\rm {R,D}}$ point can be defined as
\begin{IEEEeqnarray}{rCl}
\label{eq:Radiance}
L_{\rm R\leftarrow I_{R,D}}&=&\frac{(m+1)P_{\rm opt,S}}{2\pi} \cos^{m-1}(\theta^{\rm I_{R,D}}_{\rm {R}}) \IEEEnonumber\\
&&\times \mathbb{I}\left(|\mathbf{e}_1^{\rm T}{\rm \mathbf{S}_c\mathbf{I}_{R,D}}|\leq\frac{w_s}{2}, |\mathbf{e}_2^{\rm T}\mathbf{S}_{\rm c}\mathbf{I}_{\rm R,D}|\leq\frac{h_s}{2}\right) \IEEEnonumber\\
&& \times \mathbb{I}\left(\widehat{\mathbf{N}}_{{\rm R}}^{\rm T}\widehat{{\rm \mathbf{R}}\mathbf{D}}\geq 0\right),
\vspace{-1mm}
\end{IEEEeqnarray}
where $\theta^{\rm I_{R,D}}_{\rm {R}}$ is the irradiance angle from ${\rm {I}_{R,D}}$ to R with respect to the $z$-axis, and $\cos(\theta^{\rm I_{R,D}}_{\rm {R}})=\mathbf{e}_3^{\rm T}\widehat{\rm \mathbf{I}_{R,D}\mathbf{R}}$, where $\widehat{\rm \mathbf{I}_{R,D}\mathbf{R}}$ is computed by the law of reflection as~\cite[(2.24)]{dutre2006advanced}
\begin{IEEEeqnarray}{rCl}
\vspace{-1mm}
\widehat{\rm \mathbf{I}_{R,D}\mathbf{R}} & = & -\left(2(\widehat{\mathbf{N}}_{{\rm R}}^{\rm T}\widehat{{\rm \mathbf{R}}\mathbf{D}})\cdot \widehat{\mathbf{N}}_{{\rm R}} - \widehat{{\rm \mathbf{R}}\mathbf{D}} \right).
\end{IEEEeqnarray}
The two binary variables in \eqref{eq:Radiance} represented by $\mathbb{I}(\cdot)$ take a value of 1 when the condition inside is satisfied, and 0 otherwise. The first variable considers when the point I$_{\rm R,D}\in\mathcal{S}$, and the second variable ensures Snell's reflection law.  The point I$_{\rm R,D}$ for each R and D is computed as the incident point that the ray coming from D reflected in R has on the plane of S, computed by~\cite[(47)]{MirrorVSMetasurface}
\vspace{-1mm}
\begin{IEEEeqnarray}{rCl}
 {\rm \mathbf{I}_{R,D}}& = & \begin{bmatrix}
            \mathbf{e}_1^{\rm T}\left(\mathbf{R}+\frac{\mathbf{e}_3^{\rm T}\mathbf{RS}_{\rm c}}{\mathbf{e}_3^{\rm T}\widehat{\rm \mathbf{R} \mathbf{I}_{R,D}}}\widehat{\rm \mathbf{R} \mathbf{I}_{R,D}}\right)\\
            \mathbf{e}_2^{\rm T}\left(\mathbf{R}+\frac{\mathbf{e}_3^{\rm T}\mathbf{RS}_{\rm c}}{\mathbf{e}_3^{\rm T}\widehat{\rm \mathbf{R} \mathbf{I}_{R,D}}}\widehat{\rm \mathbf{R} \mathbf{I}_{R,D}}\right)\\
           \mathbf{e}_3^{\rm T}\mathbf{S}_c
         \end{bmatrix}.
\end{IEEEeqnarray}

The total irradiance at a point D from the whole reflective surface $\mathcal{R}$ can be written as
\vspace{-1mm}
\begin{IEEEeqnarray}{rCl}
\label{eq:TotalIrradianceAtP}
E_{\rm D}^{\mathcal{R}}&{=}&\frac{\rho_{\mathcal{R}}(m{+}1)P_{\rm opt,S}}{2\pi}\hspace{-2mm}\int_\mathcal{R} \hspace{-1mm} (\mathbf{e}_3^{\rm T}\widehat{\rm \mathbf{I}_{R,D}\mathbf{R}})^{(m{-}1)}\hspace{-0.5mm}\frac{(\mathbf{e}_3^{\rm T}\widehat{\mathbf{DR}})(\widehat{\mathbf{N}}_{\rm R}^{\rm T}\widehat{\mathbf{RD}})}{D^2_{\rm R,D}} \IEEEnonumber\vspace{-1mm}\\
&&\times \mathbb{I}\left(|\mathbf{e}_1^{\rm T}{\rm \mathbf{S}_c\mathbf{I}_{R,D}}|\leq\frac{w_s}{2}, |\mathbf{e}_2^{\rm T}\mathbf{S}_{\rm c}\mathbf{I}_{\rm R,D}|\leq\frac{h_s}{2}\right) \IEEEnonumber\vspace{-1mm}\\
&& \times \mathbb{I}\left(\widehat{\mathbf{N}}_{{\rm R}}^{\rm T}\widehat{{\rm \mathbf{R}}\mathbf{D}}\geq 0\right) d\mathcal{R}.
\end{IEEEeqnarray}

\begin{figure}[t]
     \centering
         \includegraphics[width=0.65\columnwidth]{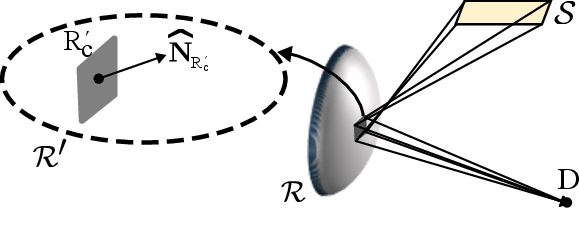}
        \caption{Representation of the small contributing area of $\mathcal{R}$.}
\label{fig:SimplifiedArea}
\vspace{-5mm}
\end{figure}

For simplification purposes, we can derive an approximation for this irradiance equation. Let us define the points $R'\in\mathcal{R}:E_{\rm D}^{R'}\neq 0$, which conform the subset $\mathcal{R'}\subset \mathcal{R}$ involving all points in $\mathcal{R}$ that contribute to D when the light is coming from any point in $\mathcal{S}$. The subset $\mathcal{R'}$ conforms a small sub-area in $\mathcal{R}$ that can be approximated as a single plane mirror of area $A_{\mathcal{R'}}$ and normal vector $\widehat{\mathbf{N}}_{\rm R'_c}$, as represented in Fig.\,\ref{fig:SimplifiedArea}, where ${\rm{R}'_c}$ represents the centroid of $\mathcal{R'}$ whose point vectors can be computed as ${\mathbf{R}'_c}=\sum_{\mathbf{R}'\in \mathcal{R}'} {\mathbf{R}'}/|\mathcal{R}'|$, and $\widehat{\mathbf{N}}_{\rm R'_c}=\left(\frac{\mathbf{R}'_\mathrm{c}\mathbf{S}_\mathrm{c}}{D_{\mathrm{R'_c},\mathrm{S_c}}} + \frac{\mathbf{R}'_\mathrm{c}\mathbf{D}}{D_{\mathrm{R'_c},\mathrm{D}}}\right)\Big/\sqrt{2+2\frac{\mathbf{R}'_\mathrm{c}\mathbf{D}^{\rm T}\mathbf{R}'_\mathrm{c}\mathbf{S}_\mathrm{c}}{D_{\mathrm{R'_c},\mathrm{S_c}}D_{\mathrm{R'_c},\mathrm{D}}}}$, as ${\rm{R}'_c}$ is located in the bisector formed by lines $\mathbf{R}'_\mathrm{c}\mathbf{S}_\mathrm{c}$ and $\mathbf{R}'_\mathrm{c}\mathbf{D}$. $|\mathcal{R}'|$ is the cardinality of set $\mathcal{R}'$, and $D_{\mathrm{R'_c},\mathrm{S_c}}$ and $D_{\mathrm{R'_c},\mathrm{D}}$ are the Euclidean distances between $\rm{R'_c}$ and $\rm{S_c}$ points, and between $\rm{R'_c}$ and D points, respectively. Once the whole reflector has been simplified to a small sub-area, we can assume that the largest dimension of the surface $\mathcal{R'}$ is much smaller than the shortest distance between a point in $\mathcal{R'}$ and a point in the source $\mathcal{S}$. Therefore every variable in \eqref{eq:TotalIrradianceAtP} depending on I$_{\rm R,D}$ and R can be replaced by S$_{\rm c}$ and ${\rm{R}'_c}$, respectively. We can follow up the large source and small reflector case approximation in \cite[Section VI.B]{MirrorVSMetasurface}, and we can write an approximation for the irradiance equation initially described in~\eqref{eq:TotalIrradianceAtP} as
\vspace{-1mm}
\begin{IEEEeqnarray}{ll}
\label{eq:SimplifiedTotalIrradianceAtP}
\widetilde{E}_{\rm D}^{\mathcal{R}}=& \IEEEnonumber\\
\frac{ \rho_{\mathcal{R}}(m{+}1)P_{\rm opt,S} A_{\mathcal{R'}} (\mathbf{e}_3^{\rm T}\widehat{\mathbf{S}_{\rm c}\mathbf{R}'_{\rm c}})^{(m-1)}(\mathbf{e}_3^{\rm T}\widehat{\mathbf{DR}'_{\rm c}})(\mathbf{N}^{\rm T}_{\rm R'_{\rm c}}\widehat{\mathbf{R}'_{\rm c}\mathbf{D}})}{2\pi D^2_{\rm R'_{\rm c},D}}. & \IEEEnonumber\vspace{-3mm}\\
\end{IEEEeqnarray}

\vspace{-0.5cm}
\section{Analytical and simulation results}\label{sec:Results}


We aim to analyze the contribution that a curved reflector may have into the signal power received by a user at a position that can be distributed along a room. We select a room with $(x,y)$ dimensions 4x4\,m. The reflector is installed in the vertical plane $y=0$, and its center is in $(x,z)=[2,1]$ coordinates. We select a mirror as the reflector surface, with a high reflection coefficient of $\rho_{\mathcal{R}}=0.99$. We consider a single light source of dimensions $w_{\rm s}=0.2$\,m and $l_{\rm s}=0.2$\,m located in the center of the ceiling, at relative coordinates $x_{\rm s}=0$\,m and $y_{\rm s}=2$\,m. The light source transmits a total optical power of $P_{\rm opt}=20$\,W with a radiation pattern defined by $\phi_{1/2}=80^\circ$. The detector can be located at any point in the horizontal plane D, and it is distributed uniformly in our simulations. The PD's responsivity and area are $\eta_{\rm pd}=0.4$\,A/W and $A_{\rm pd}=4$\,cm$^2$, respectively, and the communication bandwidth and power spectral density of the noise are assumed to be $B=1$\,MHz and $N_0=2.5\cdot 10^{-20}$\,W/Hz, respectively. All simulation parameters are summarized in Table\,\ref{tab:SystemParameters}.

\begin{table}[!t]
\caption{Simulation parameters.}
\label{tab:SystemParameters}
\centering
{\footnotesize
\setlength\tabcolsep{4.5pt} 
\begin{tabular}{llcr}
\hline
\hline
Parameter & Description & Value & Unit\\
\hline
- & $(x,y)$ room dimensions &  4x4 & [m]\\
- & Reflector center coordinates & [2; 0; 1] & [m]\\
$h_{\rm d}$ & Room ceiling height w.r.t. receiver &  2 & [m]\\
$x_{\rm s}$ & $x$-coordinate of source center (S$_{\rm c}$) &  0 & [m]\\
$y_{\rm s}$ & $y$-coordinate of source center (S$_{\rm c}$) &  2 & [m]\\
$w_{\rm s}$ & Source width &  0.2 & [m]\\
$l_{\rm s}$ & Source length &  0.2 & [m]\\
$P_{\rm opt}$  & Optical transmit power &  20 & [W]\\
$\phi_{1/2}$ & \makecell[l]{Half-power semi-angle of \\ the LED} &  80 & [deg.]\\
$\rho _{{\mathcal{R}}}$ & Reflection coefficient of mirror & 0.99 & [-]\\
$\eta_{\mathrm{pd}}$ & PD responsivity & 0.4 & [A/W]\\
$A_{\mathrm{pd}}$ & PD physical area & 4 & [cm$^2$]\\
$B$ & Communication bandwidth & 1 & [MHz]\\
$N_0$ & \makecell[l]{Power spectral density of\\ the AWGN} & $2.5{\cdot}10^{-20}$ & [W/Hz]\\
\hline
\hline
\end{tabular}}
\vspace{-4.5mm}
\end{table}

\begin{figure}[t]
     \centering
     \begin{subfigure}[b]{0.49\columnwidth}
         \centering
         \includegraphics[width=\textwidth]{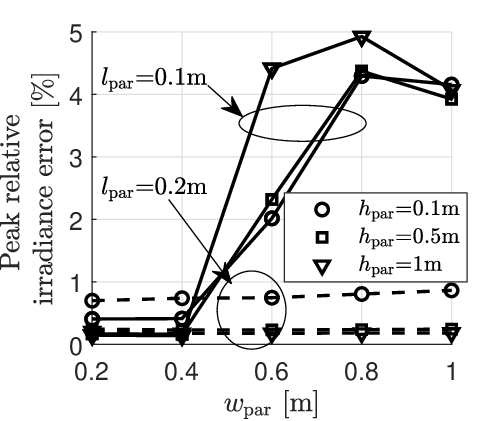}
         \caption{Paraboloid}         \label{fig:ParaboloidRelativeError}
     \end{subfigure}
          \hfill
  \begin{subfigure}[b]{0.49\columnwidth}
         \centering
         \includegraphics[width=\textwidth]{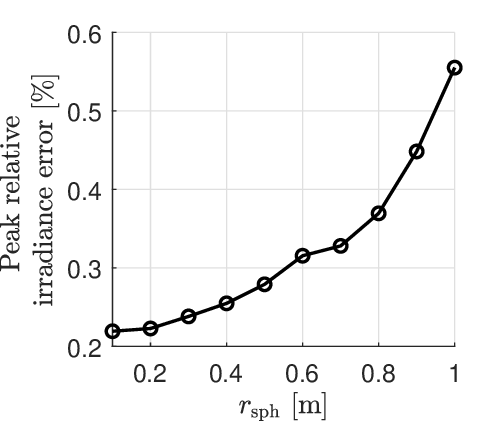}
         \caption{Sphere}         \label{fig:SphereRelativeError}
     \end{subfigure}
        \caption{Relative error between exact and approximated irradiance for multiple mirror sizes.}       \label{fig:RelativeError}
                \vspace{-5.5mm}
\end{figure}

We first evaluate the accuracy of the approximation for the irradiance equation derived in Section\,\ref{sec:IrradiancePerformance}. For this purpose, we assume that the reflector is a paraboloid with dimensions in the range of $h_{\rm par}\in[0.1,1]$\,m, $l_{\rm par}\in [0.1, 0.2]$\,m and $w_{\rm par}\in [0.2,1]$\,m, or a semi-sphere with dimensions $r_{\rm sph}\in [0.1,1]$\,m. We then evaluate the maximum (peak) relative error produced by the exact and approximated NLoS irradiance expressed in \eqref{eq:TotalIrradianceAtP} and \eqref{eq:SimplifiedTotalIrradianceAtP}, respectively. The relative error is computed as $|E_{\rm D}^{\mathcal{R}}-\widetilde{E}_{\rm D}^{\mathcal{R}}|/E_{\rm D}^{\mathcal{R}}$ for all realistic D positions, i.e., all except for those which are located at a distance from the wall $y=0$ that is lower than $l_{\rm par}$. As can be seen in Fig.\,\ref{fig:RelativeError}, the peak relative error is lower than 5\% for all configurations. It slightly increases with $w_{\rm par}$ at low curvature ($l_{\rm par}=0.1$\,m), because the shadowing effect increases as it will be seen next, and then the accuracy reduces. For the semi-sphere reflector case, it slightly increases with $r_{\rm sph}$ due to approaching the assumption limit in \eqref{eq:SimplifiedTotalIrradianceAtP} 
about the dimension of $\mathcal{R}'$ and the distance between $\mathcal{R}'$ and $\mathcal{S}$, but it is always below 0.6\%. This demonstrates the good accuracy of the approximated equation \eqref{eq:SimplifiedTotalIrradianceAtP} to compute the irradiance coming from a curved mirror. 

\begin{figure}[t]
     \centering
     \begin{subfigure}[b]{0.32\columnwidth}
         \centering
         \includegraphics[width=\textwidth]{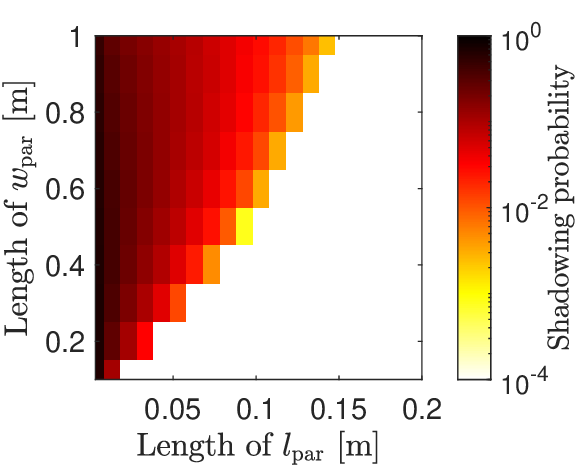}
         \caption{$h_{\rm par}=0.1$\,m}
         \label{fig:h_par01m}
     \end{subfigure}
     \hfill
     \begin{subfigure}[b]{0.32\columnwidth}
         \centering
         \includegraphics[width=\textwidth]{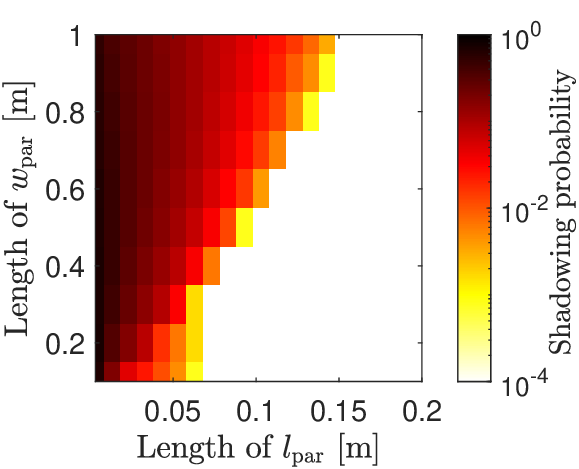}
         \caption{$h_{\rm par}=0.5$\,m}
         \label{fig:h_par05m}
     \end{subfigure}
          \hfill
  \begin{subfigure}[b]{0.32\columnwidth}
         \centering
         \includegraphics[width=\textwidth]{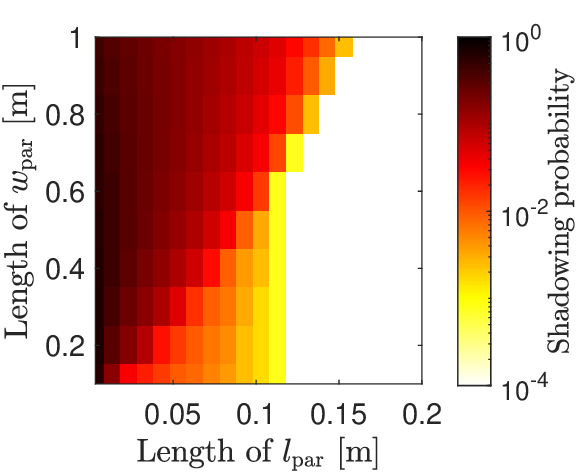}
         \caption{$h_{\rm par}=1$\,m}
         \label{fig:h_par1m}
     \end{subfigure}
        \caption{Shadowing probability for multiple combinations of $l_{\rm par}$, $w_{\rm par}$ and $h_{\rm par}$ parameters.}       \label{fig:l_par_w_par_h_parCombinations}
        \vspace{-6mm}
\end{figure}

\begin{figure}[t]
     \centering
    \begin{subfigure}[b]{0.48\columnwidth}
         \centering
         \includegraphics[width=\textwidth]{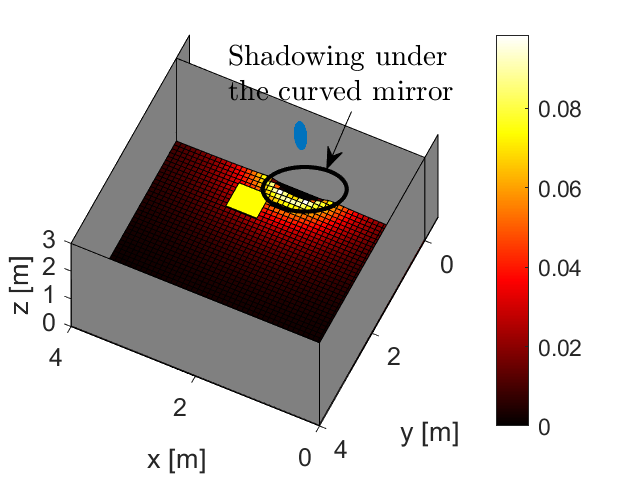}
         \caption{$h_{\rm par}=1$\,m, $l_{\rm par}=0.1$\,m and $w_{\rm par}=0.2$\,m}         \label{fig:h_par1m_l01m_w02m}
     \end{subfigure}
          \hfill
    \begin{subfigure}[b]{0.48\columnwidth}
         \centering
         \includegraphics[width=\textwidth]{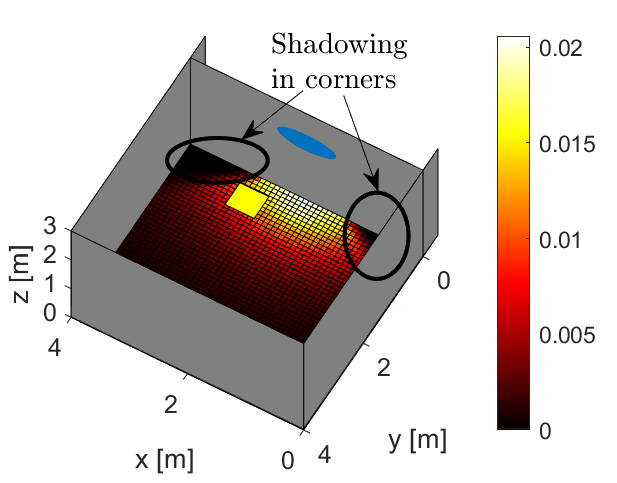}
         \caption{$h_{\rm par}=0.5$\,m, $l_{\rm par}=0.1$\,m and $w_{\rm par}=1$\,m}         \label{fig:h_par05m_l01m_w1m}
     \end{subfigure}
        \caption{3D representation of the received irradiance for two particular set of parameters for a paraboloid mirror.}       \label{fig:l_par_w_par_h_parParticularCombinations}
        \vspace{-3mm}
\end{figure}

Let us now evaluate the performance of the curved mirror when it has different dimensions. We analyze the shadowing probability defined by \eqref{eq:ShadowingProb}. Fig.\,\ref{fig:l_par_w_par_h_parCombinations} shows the shadowing probability for multiple combinations of $l_{\rm par}$, $w_{\rm par}$ and $h_{\rm par}$ when using a paraboloid-shaped mirror reflector. We consider that $h_{\rm par}$ may take values of 0.1\,m, 0.5\,m, or 1\,m for a small, medium, or large mirror, respectively. Then $w_{\rm par}$ can go from 0.1\,m to 1\,m, and the curvature of the mirror defined by the $l_{\rm par}$ parameter can go from 0\,m (plane mirror) to 0.2\,m. Results show that the larger the height of the mirror denoted by $h_{\rm par}$, the higher the shadowing probability is due to a curvature decrease, which introduces some shadowing just below the reflector as shown in Fig.\,\ref{fig:h_par1m_l01m_w02m}. A similar phenomenon happens when $w_{\rm par}$ increases for the same $h_{\rm par}$ and $l_{\rm par}$ values, which is shown in Fig.\,\ref{fig:h_par05m_l01m_w1m}, where we can see that a lower curvature introduces shadowing in the room corners. However, for the same $h_{\rm par}$ and $w_{\rm par}$ values, a larger curvature ($l_{\rm par}$) makes the shadowing probability reduce, since the visibility at the room edges increases. This gives the insight that, given a mirror size defined by $h_{\rm par}$ and $w_{\rm par}$, there is a minimum $l_{\rm par}$ so that the shadowing probability equals zero. 
Although a graph representation is not included in this paper, note that, when using a semi-sphere-shaped mirror reflector, the shadowing probability is zero due to its perfect curvature.

    \begin{figure}[t]
     \centering
         \includegraphics[width=0.9\columnwidth]{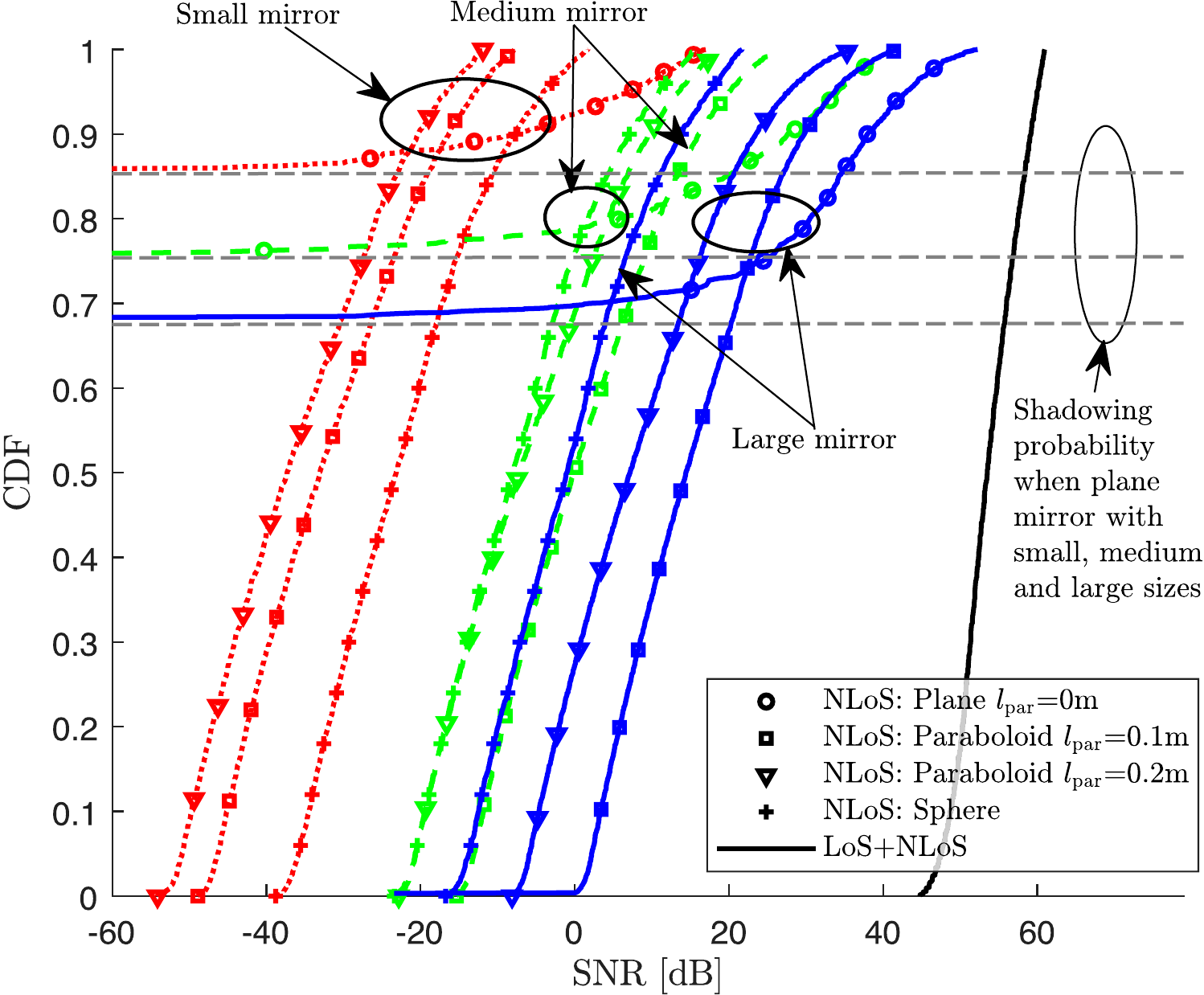}
        \caption{Contribution from NLoS in case of LoS blockage. CDF of optical received power under different scenarios.}
        \label{fig:LoSNLoSSNR}
        \vspace{-6mm}
\end{figure}

\begin{table}[!t]
\caption{Dimensions of paraboloid and semi-sphere mirrors.}
\label{tab:DimensionsMirrors}
\centering
 \setlength\tabcolsep{3pt} 
\begin{tabular}{|l|l|l|l|}
\hline
\textbf{Size} & \textbf{Area [m$^2$]} & \textbf{\makecell[l]{Dimensions in \\ paraboloid case [m]}} & \textbf{\makecell[l]{Dimensions in \\ semi-sphere case [m]}}\\
\hline
Small & 0.0314 &  $w_{\rm par}=0.4$; $h_{\rm par}=0.1$  & $r_{\rm sph}=0.1$ \\
Medium & 0.1571 & $w_{\rm par}=0.4$; $h_{\rm par}=0.5$ & $r_{\rm sph}=0.2236$\\
Large & 0.3142 & $w_{\rm par}=0.4$; $h_{\rm par}=1$ & $r_{\rm sph}=0.3162$\\
\hline
\end{tabular}
\vspace{-6mm}
\end{table}

Finally, we evaluate the SNR defined in \eqref{eq:SNR} when the LoS is blocked (NLoS case) or not (LoS + NLoS case). For the NLoS case, we analyze a number of cases that are grouped by the area occupied by the mirror on the wall. We analyze three areas according to the paraboloid and semi-sphere dimensions: small size (area $= 0.0314$ m$^2$), medium size (area $=0.1571$ m$^2$), and large size (area $= 0.3142$ m$^2$). Their dimensions are included in Table\,\ref{tab:DimensionsMirrors}, and 
the results are depicted in Fig.\,\ref{fig:LoSNLoSSNR}. As expected, the larger the size, the larger the SNR obtained. Plane mirrors provide a shadowing floor, i.e., locations with a SNR $=-\infty$\,dB, with probability values of 68\%, 76\% and 86\% for large, medium and small mirrors. However, the use of curved mirrors reduces the shadowing probability to zero, as seen in previous results and also in Fig.\,\ref{fig:LoSNLoSSNR}. In terms of the SNR distribution, the best curvature $l_{\rm par}$ is an intermediate value of 0.1\,m, as a very low $l_{\rm par}$ value (plane mirror) increases the shadowing probability, and very large $l_{\rm par}$ values distribute the light to the room edges to a very large extent. Thus, there is a balance in the selection of the $l_{\rm par}$ value that reduces the shadowing probability while maintaining a good light distribution. When comparing the paraboloid and semi-sphere mirror shapes, the paraboloid provides better performance when the size of the mirror is medium or large. However, for small area sizes, it is more convenient to use a semi-sphere as it will distribute reflections in a better way for the evaluated configurations. Note that the dimensions of the convex mirror determine its curvature, which plays a key role in the light distribution as shown in Fig.\,\ref{fig:l_par_w_par_h_parCombinations} and Fig.\,\ref{fig:l_par_w_par_h_parParticularCombinations}, and they must be carefully configured for the SNR improvement.

The results show that the LoS link provides much larger SNR values than NLoS links. NLoS contributions may be insignificant when the LoS link exists, but they are extremely important in the case of a LoS link blockage. That is, curved mirrors allow the user to have communication even when the LoS link is blocked, regardless of its location, as the SNR values provided by the NLoS link are good enough to invoke modulation and coding schemes~\cite{RatevsSNR}.

\vspace{-0.1cm}
\section{Conclusion} \label{sec:Conclusions}
\vspace{-0.1cm}
This paper investigated the contribution of curved mirrors to improve the VLC coverage. We derived equations for the irradiance received from a curved mirror with a paraboloid or semi-sphere shape, and we provided an approximation that matches the irradiance accurately. We studied the influence of multiple mirror dimensions in the shadowing probability, and results reveal that curved mirrors may offer a zero shadowing probability. 
Finally, we studied the SNR contribution coming from curved mirrors, and results show that curved mirrors are interesting static reflective surfaces to provide connectivity to the users located along the room when they experience a LoS link blockage. As future work, experimental research must be done to validate the performance of curved mirrors in VLC scenarios relying on NLOS links. Additionally, we can consider small-sized curved mirrors provided with mobility for a dynamic adaptation to minimize the intrusiveness of such mirrors type. Then, we could optimize the mirrors' curvature and placement for best performance.

\vspace{-0.2cm}



\bibliography{./references}
\bibliographystyle{IEEEtran}

\end{document}